\documentclass[12pt]{article}
\usepackage{epsfig,latexsym}

\newcommand{\siml} {\,{}^<_{\sim}\,}
\newcommand{\simg} {\,{}^>_{\sim}\,}
\begin{document}
\begin{titlepage}                            
\enlargethispage{3.cm}

\title{Cosmological Deuterium Production in Non-Standard Scenarios}
\author{Karsten Jedamzik \\ %\thanks{e-mail: jedamzik@mpa-garching.mpg.de} \\
Max--Planck--Institut\\
f\"ur Astrophysik \\
Karl--Schwarzschild--Str. 1\\
85741 Garching, Germany\\
and\\
Physique Math\'ematique et Th\'eorique, \\
Universit\'e Montpellier II, \\
34095 Montpellier Cedex 5, France}
%Physique Math\'ematique et Th\'eorique, UMR No 5825--CNRS, \\
%Universit\'e Montpellier II, F--34095 Montpellier Cedex 5, France}            

\maketitle
\begin{abstract}
It is widely believed that the cosmic baryon density may 
be obtained by inferring deuterium abundances
in low-metallicity quasar absorption line systems.
The implicit assumptions which enter this argument are critically assessed.
In particular, the production of deuterium in non-standard Big Bang 
nucleosynthesis scenarios, the putative production of deuterium in
astrophysical environments, and the possible destruction of deuterium via
non-standard chemical evolution are discussed.  
\end{abstract}
\end{titlepage}

\small\normalsize                  

\section{Introduction}
It was realized about twenty-five years ago 
(Epstein {\it et al.} 76, and references therein) that the deuterium isotope
may play a very special role in cosmology as it seems extraordinarily 
difficult to produce it in abundance in environments other than that of
a Big Bang nucleosynthesis (BBN) freeze-out process. 
The importance of the deuterium isotope to cosmology 
has increased with
the recently acquired ability of inferring comparatively precise 
D/H abundance ratios in high redshift quasar absorption line 
systems (QAS). It is now widely believed that deuterium 
is produced {\it only} during the BBN process and that the BBN process occurred
in its standard version (to be specified below). In this case there exists one
unique primordial D abundance independent of spatial location, 
and by observing 
D/H ratios in any one high-redshift system at low metallicity 
one may immediately infer the cosmic baryon density. The subject of this
talk is to critically assess the viability of parts of this line of 
argument. In this spirit I discuss D production in some non-standard
BBN scenarios in Section 2 and the possible production of D in
astrophysical environments in Section 3. Section 4 contains a few comments
on the possibility of significant stellar destruction 
of D in high-redshift, low-metallicity QAS whereas
conclusions are drawn in Section 5. Due to the breadth of the subject I
apologize in advance for not being able to address all the work which  
touches on this subject.

\section{Production of deuterium in non-standard BBN models}

Making definite theoretical predictions for the production of 
light-element abundances during a BBN freeze-out process involves making
a number of assumptions. In the standard model of BBN these assumptions are, 
among others, a uniform baryon density, the existence of 
three, light, stable, and nondegenerate active neutrino species, as well as a
a \lq\lq clean\rq\rq\ universe during the BBN era (no relic decaying 
particles, no pockets of antimatter, no defects, etc.). Furthermore, the 
validity of classical relativity and
non-varying physical constants (e.g. fine structure constant 
$\alpha$, ...) during BBN is assumed. There exists a vast literature
(cf. Malaney \& Mathews 93, Sarkar 96 for reviews) of 
modified BBN abundance predictions when any one of these assumptions is 
relaxed. It is typically not difficult to produce D in abundance
as required by observations in many of these scenarios. Rather, non-standard
scenarios are distinguished from standard ones by a different resulting 
abundance pattern in the synthesized elements D, $^3$He, $^4$He, and $^7$Li 
at varying cosmic baryon density. They may be often excluded by the 
combination of observational limits on all primordial light-elements 
abundances rather than the limit of D by itself. In the following I will 
briefly discuss three such non-standard BBN scenarios.

\subsection{inhomogeneous BBN}

Inhomogeneities in the baryon number may result from non-equilibrium
processes in the early universe, for example, occurring during a
putative first-order quantum-chromodynamics (QCD) phase transition.
Due to the smallness of the cosmological horizon during the QCD epoch
such scenarios would typically lead to small-scale inhomogeneity 
with the baryonic mass of individual fluctuations extremely small on an
astrophysical reference scale. If the mass of individual fluctuations exceeds
$M_b\simg 10^{-21}M_{\odot}$ dissipative processes in the early Universe
do not erase the inhomogeneities before the onset of BBN (at cosmic 
temperature $\sim 1\,$MeV). A large number of authors have given 
abundance predictions for such inhomogeneous scenarios 
(cf. Kurki-Suonio \& Sihvola 01, Jedamzik \& Rehm 01, and
references therein). Compared to a standard BBN 
at the same $\Omega_b$, baryon inhomogeneous models may result in the
production of more D and less $^4$He.
Considering observational limits on D and $^4$He only, larger
$\Omega_bh^2 \siml 0.05$ 
than in a standard BBN scenario ($\Omega_bh^2 \approx 0.02$; Burles 
{\it et al.} 01, O'Meara {\it et al.} 01) 
are observationally allowed. Here $\Omega_b$ is the fractional 
contribution of baryons to the 
critical density and $h$ is the Hubble constant in units of 100$\,$km 
s$^{-1}$Mpc$^{-1}$. Nevertheless, $^7$Li/H ratios at least as large as 
$5\times 10^{-10}$ - $10^{-9}$ are typical in these models and seem in
conflict with the observationally inferred primordial $^7$Li abundance.
On the other hand, even a standard BBN scenario at the preferred  
$\Omega_bh^2 \approx 0.02$ (from observations of D/H in QAS)
predicts $^7$Li/H $\approx 3.8\times 10^{-10}$ in excess of the
observationally inferred $1.2\times 10^{-10} - 1.7\times 10^{-10}$ from
low-metallicity Pop II halo stars
(cf. Bonifacio \& Molaro 97, Ryan {\it et al.} 00). Thus, most inhomogeneous
BBN models are observationally disfavored with the loophole of an as yet
not complete understanding of possible $^7$Li astration in Pop II halo stars.

One may also envision inhomogeneity in the baryons on scales much larger,
i.e. $M_b\simg 10^{5}M_{\odot}$. Such large-scale inhomogeneity could be 
in principle generated during an inflationary period with a subsequent 
inhomogeneous baryogenesis process triggered by the inhomogeneities produced
during inflation. The difference to a scenario of small-scale inhomogeneity
is the importance of gravity, i.e. individual overdense fluctuations on scales
larger than the baryonic Jeans mass ($M_J^b\approx 10^{5}M_{\odot}$) collapse
at comparatively high redshift $z\approx 10^3$ and may thereby ``hide'' their
elevated primordial $^4$He and $^7$Li BBN yields. Such collapsing regions
would either lock up their BBN yields in compact objects, or alternatively, 
undergo significant early star formation and metal enrichment 
such that they would never be employed to infer 
primordial abundance yields (which are typically
inferred from low-metallicity environments). In this case the
observed ``primordial'' abundances are from the BBN yields in 
underdense regions. It is interesting to note that such scenarios predict
a primordial dispersion in D/H (Jedamzik \& Fuller 1995) such that there exists
no unique primordial D/H abundance. Note that, though claimed in the
literature (Copi {\it et al.} 97) the author strongly disagrees on the conclusion
that primordial D/H fluctuations automatically predict large, and 
observationally 
disallowed, fluctuations in the anisotropy of the cosmic microwave background
radiation (CMBR). 
Rather, probably the strongest argument against such scenarios
is that there has to exist a cutoff in the fluctuation power 
for fluctuations with baryonic mass scale below the Jeans mass, 
as otherwise such 
non-collapsing sub-Jeans regions yield to an overproduction of $^4$He and 
$^7$Li (Kurki-Suonio {\it et al.} 97). Furthermore, such scenarios may (possibly) 
be inconsistent with the cosmic microwave background radiation due 
to early reionization and a general {\it suppression} of the CMBR (Hu 
\& Sugiyama 94).

\subsection{neutrino degeneracy}

Cosmic lepton number may be associated with a chemical potential. 
In this case an asymmetry between neutrino- and antineutrino-
number exists which, if large enough, affects the expansion rate during 
BBN and the n/p-ratio shortly before weak freeze-out. Such neutrino
degenerate models have three additional free parameters as compared to
a standard BBN, in particular, 
$\xi_{e} = \mu_e/T_{\nu}$, $\xi_{\mu}$, and $\xi_{\tau}$, where
the $\mu$'s are chemical potentials for the three families of leptons
($e$, $\mu$, and $\tau$) and $T_{\nu}$ indicates neutrino temperature.
It has been known for some time (e.g. Kang \& Steigman 92; and references therein)
that with the right ``choice'' of asymmetry parameters, light-element
abundance yields may obey observational constraints for essentially
arbitrary $\Omega_b$. For given D/H$\,\approx 3\times 10^{-5}$ the
$^4$He abundance may be somewhat
lower and $^7$Li is almost unchanged as compared to a standard BBN.
Nevertheless, in order to achieve consistency with the observations
over a wide range of $\Omega_b$ not only have the chemical potentials to
be orders of magnitude larger than that for baryon number, but chosen
``asymmetrically'', i.e. $\mu_e\sim 1$ and $\mu_{\mu,\tau}\sim 10$.
Such an asymmetry may not necessarily survive the effects of neutrino
oscillations in the very early Universe, particularly for neutrino 
mixing parameters consistent with the atmospheric neutrino anamoly and 
the large-angle mixing solution of the solar neutrino problem 
(Dolgov {\it et al.} 02). There are additional constraints on neutrino degenerate
BBN from observations of the CMBR anisotropies.
Though first results of such observations by Boomerang and Maxima 
favored $\Omega_bh^2 \simg 0.02$, possibly explained by neutrino degenerate BBN
(Orito {\it et al.} 00), more recent data seems to be consistent 
with $\Omega_bh^2 \approx 0.02$. This recent data may also be used
to derive more stringent constraints on cosmic neutrino degeneracy, possibly
already ruling out most large $\Omega_b$ degenerate BBN scenarios
(Hansen {\it et al.} 02).

Depending on the details of leptogenesis, it is possible that
lepton chemical potentials are spatially varying. Inspired by varying
claimed D/H abundances in low-metallicity QAS a number of years ago
Dolgov \& Pagel 99 have investigated such a putative scenario. Similarly
to the case of large-scale baryon inhomogeneity, they found that such
scenarios may lead to spatially varying primordial D abundances and are
currently not necessarily ruled out by CMBR anisotropy measurements.
Though both of these scenarios may seem exotic, they illustrate the
importance of an unbiased stance of observers (and theoreticians) 
towards inferring and interpreting
D/H ratios in high redshift QAS (cf. also Webb {\it et al.} 97). 
In particular, it is
conceivable, though not necessarily likely, that there does not 
exist one unique primordial D abundance.

\subsection{radiative decay of particles after BBN}

Deuterium may be destroyed (or produced) shortly 
after a standard BBN epoch
by the radiative decay of a long-lived relic particle
(Lindley 80, Audouze {\it et al. 85}). In general, injection of energetic,
electromagnetically, interacting particles by, for example, the
decay of relic particles (i.e. gravitinos or moduli) or the evaporation of
primordial black holes, leads to an electromagnetic
cascade on the cosmic microwave background radiation. 
This rapid cascade results in the production of $\gamma$-rays 
with energies below
$E_{\gamma} \siml E_{\gamma}^{th}\approx m_e^2/2E_{{\rm CMBR}}$ and of spectrum
and number only dependent on the injected electromagnetic energy but not
the details of the injection mechanism. In the above $m_e$ is electron mass
and $E_{{\rm CMBR}}$ is the typical energy of a CMBR photon
at the injection redshift.
For cosmic times $t \simg 10^4\,$s the threshold energy 
$E_{\gamma}^{th}$ is above the binding energy of the D isotope 
$E_{\rm D}^b\approx 2.2\,$MeV such that deuterium
photodisintegration may result. Such ``decaying particle'' scenarios result
in a diminished deuterium abundance compared to a standard BBN at the same
$\Omega_b$, whereas the $^4$He  and $^7$Li abundances stay virtually
unchanged. The required relic energy density $\rho_{relic}$ 
to destroy the BBN synthesized deuterium by a factor of order unity is
$\rho_{relic}/\rho_{b} \sim 10 - 10^3$, where $\rho_b$ is the baryonic 
mass density. The attraction of such models is the possibility
of obtaining low primordial D {\it and} low primordial $^4$He abundances
at the same time (Holtmann {\it et al.} 96). A standard BBN with D/H $\approx
3\times 10^{-5}$ (at $\Omega_bh^2\approx 0.02$) predicts a $^4$He mass 
fraction of $Y_p\approx 0.247$. It is currently
controversial if observational limits on the primordial $^4$He abundance
permit a $Y_p$ as large. In contrast,``decaying particle''
scenarios at lower $\Omega_bh^2$ may accommodate both, low D and low 
$^4$He. Nevertheless, though it seems not difficult to produce the
appropriate abundance of relics during the evolution of the early 
Universe, their decay time has to be somewhat tuned. 
This is because $E_{\gamma}^{th}$ at
$t \simg 10^6\,$s, exceeds $20\,$MeV, sufficient to photodisintegrate not
only D but also $^4$He. Due to the larger abundance of $^4$He the latter
reaction dominates and leads to production of copious amounts of $^3$He, D, 
and $^6$Li (cf. Section 3.1). 
Thus, deuterium may generally only be destroyed for
electromagnetic energy injection in the time interval 
$10^4\,{\rm s}\siml t \siml 10^6\,$s.

\section{``Exotic'' production sites of deuterium}

It is common belief that there exist {\it no} deuterium production
sites other than the BBN and that deuterium is only subject to 
stellar destruction. This statement not necessarily holds true though. 
Rather, there are no believable production sites which
may produce deuterium in cosmological abundance without overproducing
other isotopes, in particular, $^3$He, $^6$Li, and $^7$Li 
(Epstein {\it et al.} 76).
Nevertheless, in the absence of (local) abundance determinations of these
complimentary 
light-element isotopes (e.g. in QAS) it is, in principle, possible that
deuterium in a given system has been locally enhanced by ``exotic''
mechanisms. With the advent of accurate deuterium determinations in QAS 
the study of such ``exotic'' processes has become feasible, and 
represents an opportunity. There exist principally three straightforward
ways to produce deuterium: by the photodisintegration of $^4$He, by the
spallation of $^4$He, as well as by the production of free neutrons and
subsequent fusion of these neutrons on protons. In the following subsection
each of these possibilities will be discussed.

\subsection{production by $^4$He photodisintegration}

Deuterium may be produced by the photodisintegration processes
$^4$He($\gamma$,np)D and $^4$He($\gamma$,D)D whenever there exist 
$\gamma$-rays of energy exceeding
$E_{\gamma}\simg 23\,$MeV. This may occur cosmologically by the 
electromagnetic decay of relics of the early universe (cf. Section 2.3) after 
BBN (e.g. gravitinos; Audouze {\it et al.} 1985),
or locally, close to an intense $\gamma$-ray source (e.g. accreting
black holes Gnedin \& Ostriker 92). Nevertheless, the processes 
$^4$He($\gamma$,p)$^3$H and $^4$He($\gamma$,n)$^3$He possible for 
$E_{\gamma}\simg 20\,$MeV 
are typically $\sim 10$ times more likely. Thus $^4$He-photodisintegration  
yields an unacceptable large 
ratio $^3$He/D $\sim 10$ (Protheroe {\it et al.} 95, Sigl {\it et al.} 95),
unless the $\gamma$-ray source is very hard with emission dominated
above $E_{\gamma}\simg 100\,$MeV. In this latter case production ratios
of $^3$He/D $\sim 1$ are possible. Nevertheless, quite independently
of the hardness of the $\gamma$-ray source, significant overproduction
of $^6$Li is almost guaranteed (Jedamzik 00). This is due to the produced
$^3$He being nonthermal, partaking in fusion reactions on $^4$He 
to form $^6$Li. For example, a cosmological production of D/H $\approx
3\times 10^{-5}$ by the decay of a relic after BBN is concomitant with
the production of $^6$Li/H-ratios of the order of $10^{-9}$. 
This should be compared to the $^6$Li/H $\approx 7\times 10^{-12}$ 
as observed in low-metallicity halo stars (Cayrel {\it et al.} 99).
Concerning the $^3$He/D ratio there exists a limit (under standard
assumptions) derived from its presolar value 
$({\rm ^3He/D})_{\odot}\approx 1$.  
Deuterium is only destroyed during stellar processing whereas $^3$He may be
destroyed or produced in stars. This implies that the $({\rm ^3He/D})$
may only increase with time and immediately rules out any processes
which yield large $^3$He/D-ratios (such as decaying relics after BBN) as
the production mechanism for the deuterium inferred to exist at the formation
of the solar system (Sigl {\it et al.} 95).

Nevertheless, in the absence of detailed abundance determinations 
of various  isotopes (e.g. $^3$He and $^6$Li) in the same ``astrophysical''
object, one may not exclude the possibility that, for example,
a particular QAS has been enhanced in D by $^4$He photodisintegration.  
Suppose there exists a putative population of powerful $\gamma$-ray
sources, bursting at redshift $z_b$. The probability that a line of
sight towards a quasar passes through the zone of significant
deuterium enrichment by $^4$He photodisintegration may be written as
(Jedamzik \& Fuller 97):
$P \sim 10^{-9} (1+z_b) (j_{\gamma}(z=0,E_{th}/(1+z_b))/10^{-5}
{\rm MeV^{-1}cm^{-2}s^{-1}sr^{-1}})$, and is found
very small unless, the bursts happen at large $z_b$. In the above
expression $j_{\gamma}(z=0,E_{th}/(1+z_b))$ is the specific x/$\gamma$-ray
intensity at the present epoch determined at the energy $E_{th}/(1+z_b))$
and $10^{-5} {\rm MeV^{-1}cm^{-2}s^{-1}sr^{-1}}$ is the approximate
present specific intensity at $E_{\gamma}\approx 20\,$MeV. It may be
concluded that local, or global, D-enrichment by $^4$He photodisintegration
seems unlikely. 

\subsection{production by $^4$He spallation}

Deuterium production may also result from the interactions of a cosmic
ray population (energetic p, $^4$He with $\sim$ a few 100 MeV) with
interstellar matter, by the spallation processes
p($^4$He,$^3$He)D and p($^4$He,p)2D. Nevertheless, efficient production
can not result from (locally) known cosmic ray populations.
The spallation production of D is accompanied by
production of $^3$He as well as 
$^6$Li and $^7$Li via fusion of energetic cosmic ray $^4$He on 
interstellar $^4$He, i.e. $^4$He + $^4$He $\to$ $^6$Li, $^7$Li, and $^7$Be.
Unlike in the case of D production via $^4$He photodisintegration,
resulting $^3$He/D ratios in spallation may be close to 
unity, at least for putative (low-energy) cosmic ray populations 
with a few 100 MeV energy per particle (Famiano {\it et al.} 01). 
Nevertheless, the 
Li-overproduction problem is even more
severe in scenarios which envision to produce significant amounts of D 
by spallation
processes. Overproduction of $^7$Li by orders of magnitude relative to
the $^7$Li/H observed in low-metallicity halo stars belonging to the 
Spite-plateau (i.e. typical production of $^7$Li/H
$\sim 10^{-7}$ for D/H $\sim 10^{-4}$), as well
as overproduction of $^6$Li precludes spallation processes from explaining
the origin of cosmological D and $^3$He (Epstein {\it et al.} 76).
Moreover, in the presence of CNO nuclei spallation processes would also
yield to significant production of $^9$Be, $^{10}$B, and $^{11}$B
when CNO nuclei are present. Such processes are, in fact, believed
to be the main source of these latter isotopes.
In any case, local D enhancement due to spallation processes having occurred
in an object are conceivable, as long
as other light elements are not observed (or do not violate observational
constraints). Recently, Famiano {\it et al.} 01 have proposed 
that the interactions of jets originating from active galactic nuclei with
interstellar clouds may result in efficient local D enhancement. 
It is not known how common such processes are, 
and if such clouds would appear as objects where
inference of primordial D/H is attempted, i.e. 
low-metallicity high redshift Lyman-limit- or damped Lyman-$\alpha$- systems. 
It illustrates, however, that in interpreting inferred D/H abundances in
QAS one should act with care as to not overcommit to the bias that the only
production site for D is BBN.

\subsection{production in an intense neutrino flux}

Deuterium may be produced by creating free neutrons which subsequently
capture on protons, i.e. p(n,$\gamma$)D. Neutron production may occur via
the weak interaction $\bar{\nu}_e + p \to n + e^+$ in the
presence of an intense anti-electron neutrino flux. In order to form 
deuterium, capture of the produced neutrons on protons has to occur before 
the neutrons decay $n\to p + e^- + \bar{\nu}_e$ with a half life of
10.26 min. Furthermore, D is very fragile and easily fuses to $^3$He at
low temperatures, i.e. D(p,$\gamma$)$^3$He, such that the environment in which
this process could successfully occur
has to be relatively cool or rapidly cooling down. 
The conditions for efficient production of D via such a mechanism, in
particular, existence of an intense neutrino flux, 
rapid neutron capture, and
inefficient D-burning, may be naturally met in supermassive stars of 
$\sim 10^5M_{\odot}$, towards the end of their evolution and shortly
before black hole formation (Woosley 77, Fuller \& Shi 97). 
In fact, Woosley 77
proposed this mechanism for the origin of the bulk of the
cosmologically observed deuterium. This, nevertheless, seems unlikely, 
as it entails a large fraction of baryons to be locked up in massive 
black holes at the present epoch. Other than leaving black hole remnants, a
scenario where D is locally enhanced by supermassive stars is clean
from a nucleosynthetic point of view. This is because no heavier elements
than $^4$He are produced in an early population of $\sim 10^5M_{\odot}$ stars.
It seems, however, uncertain what fraction of the
deuterium enriched zones may get expelled into the interstellar medium and
what fraction ends in the black hole. 
If a large fraction gets expelled, 
then these ejecta may appear as deuterium enhanced
Lyman-limit systems and are found with optical depth towards quasars of
$\tau \sim 100\, \Omega_{sms} (1+z)^3 (R/1{\rm kpc})^2 M_5^{-1}$
(Fuller \& Shi 97). In the above, $\Omega_{sms}$ is the fractional contribution
of baryons to the critical density which has been cycled through
supermassive stars at redshift $z$, $R$ is the expansion radius of the 
deuterium enriched envelope, and $M_5$ is the stellar mass in units of
$10^5M_{\odot}$. It may be seen that even for fairly small 
$\Omega_{sms}\sim 0.01\%$ the probability of any given line of sight towards
a high redshift quasar to pass through such a system is appreciable.
In view of the current lack of understanding of the origin of supermassive
black holes at the center of galaxies, the above drawn possibility has, 
in fact, to be taken seriously and deserves further study.

\section{A simple argument against deuterium depletion at high redshift?}

Deuterium is easily destroyed in stars due to nuclear burning. In order
to infer a primordial deuterium abundance it is thus
important to constrain the degree of stellar D
depletion which may have occurred in high redshift QAS. 
This is often done by inferring the abundance of one, or two, heavier
isotopes (e.g. C and Si) and under the assumption of ``normal'', average 
star formation having taken place in the QAS.   
Average stellar destruction of deuterium at high redshift is then expected
to be small, as inferred from the chemical abundances 
of damped Lyman-$\alpha$ systems
and under the assumption of an ordinary 
stellar initial mass function (IMF) as observed
in our local neighborhood.
Nevertheless, due to their distance not much is known about the QAS used to 
infer D abundances. It is thus conceivable that 
stochastic or anomalous stellar processing 
may result in significant destruction of deuterium
in individual QAS. This could occur, for example, when the gas which is
metal-enriched by the ejecta of massive stars is preferentially 
expelled (outflows), or
when stars in only a very small mass range 
form (peaked IMF), where production
of heavy isotopes is inefficient. 
A simple argument against such possibility was put forward by
Jedamzik \& Fuller 97.
The universe is only $\approx 2\,(3)\,$Gyr old at redshift $z \approx 3 (2)\,$.
Stars with $M \approx 1.5 (1.35)\, M_{\odot}$ live about $2\, (3)\,$Gyr before
they may return their D-depleted ejecta into the ISM. 
It is (see below, however) 
thought that stars of $\sim 1.5 (2) - 4\,M_{\odot}$ are efficient
$^{12}$C producers, stars of $\sim 4 - 8\,M_{\odot}$ are efficient $^{14}$N
producers, and stars with $M > 8\,M_{\odot}$ are producing $^{28}$Si and
$^{56}$Fe. This would imply that local stellar deuterium depletion 
has to be accompanied by enhancement of heavy elements at high redshift.
Stars which have negligible production of isotopes heavier
than $^4$He are either very low in mass $\siml 1.35\, M_{\odot}$ and have not 
had time to return their D depleted ejecta into the interstellar medium
at high redshift, or supermassive (cf. Section 3.3). Note that this constrain
would be independent of the loophole of outflows since the {\it same} stars
which deplete deuterium also produce the ``metals''.
However, recent investigation has cast doubt on the belief that 
intermediate mass zero-metallicity stars actually release their synthesized
C and N in a ``dredge-up'' event (cf. Fujimoto {\it et al.} 00, Marigo 
{\it et al.} 01, Chieffi {\it et al.} 01). This has tempted Fields {\it et al.}
(these proceedings) to speculate on the possibility of significant 
stellar D depletion in QAS and the concomitant production of an old white
dwarf population. Clearly, the subject of metal enrichment by 
zero-metallicity stars deserves further study.
  
\section{Conclusions}
The bulk of the cosmological deuterium is believed to originate
from the hot Big Bang. Decades of research have not revealed another viable
alternative for its origin; alternatives are usually ruled out
by overproduction of other isotopes (i.e. $^3$He, $^6$Li, and $^7$Li).
Moreover, deuterium production may well have occurred in
the simplest (standard) version of BBN, as there is currently
no compelling evidence for a non-standard BBN scenario to be required.
Nevertheless, I have attempted to illustrate that local, so far unknown, 
sources (sinks) of deuterium may exist. They imply the possibility 
of the occasional deuterium enhanced (or depleted) 
quasar absorption lines system.
Particularly promising in this context may be the production of D by a 
generation of supermassive stars. 
Abundance determination of deuterium at extragalactic distances
may therefore {\it not only} be used to derive the primordial deuterium
abundance, but also, to discover or constrain other astrophysical processes
in the Universe.

%\begin{thebibliography}{20}
%\begin{thebibliography}
\vskip 0.8in

\vskip 0.1in
\noindent
Audouze, J., Lindley, D., Silk, J., 1985.
Big bang photosynthesis and pregalactic nucleosynthesis of light elements.
Astrophys. J. 293, L53-L57.

\vskip 0.1in
\noindent
Bonifacio, P., Molaro, P., 1997. The primordial lithium abundance.
Mon. Not. Roy. Astron. Soc. 285, 847.

\vskip 0.1in
\noindent
Burles, B., Nollett, K. M., Turner, M. S. 2001.
Big-Bang nucleosynthesis predictions for precision cosmology.
Astrophys. J. 552, L1-L6.

\vskip 0.1in
\noindent
Cayrel, R., Spite, M., Spite, F., Vangioni-Flam, E., Cass\'e, M., Audouze, J., 1999.
New high S/N observations of the Li-6/Li-7 blend in HD 84937 and two 
other metal-poor stars.
Astron. and Astrophys., 343, 923-932.

\vskip 0.1in
\noindent
Chieffi, A., Dominguez, I., Limongi, M., Straniero, O., 2001.
Evolution and nucleosynthesis of zero metal intermediate mass stars.
Astrophys. J. submitted, astro-ph/0103104.

\vskip 0.1in
\noindent
Copi, C. J., Olive, K. A., Schramm, D. N., 1997.
Implications of a primordial origin for the dispersion in D/H 
in quasar absorption systems. astro-ph/9606156

\vskip 0.1in
\noindent
Dolgov, A. D., Pagel B. E. J., 1998. Varying lepton chemical
potentials and spatial variation of primordial deuterium at high z.
New Astron. 4, 223-230.

\vskip 0.1in
\noindent
Dolgov, A. D., Hansen, S. H., Pastor, S., Petcov, S. T., Raffelt, G. G., 
Semikoz, D. V., 2002. Cosmological bounds on neutrino degeneracy improved
by flavor oscillations. hep-ph/0201287.

\vskip 0.1in
\noindent
Epstein, R. I., Lattimer, J. M., Schramm, D. N., 1976.
Origin of deuterium. Nature, 263, 198-202. 

\vskip 0.1in
\noindent
Famiano, M, Vandegriff, J., Boyd, R. N., Kajino, T., Osmer,
P., 2001. Production of $^2$H, $^3$He, and $^7$Li from interactions between
jets and clouds. Astrophysical J. 547, L21-L24.

\vskip 0.1in
\noindent
Fujimoto, M. Y., Ikeda, Y., Iben, I. Jr., 2000.
The origin of extremely metal-poor carbon stars and the search for population III.
Astrophysical J., 529, L25-L28.

\vskip 0.1in
\noindent
Fuller, G. M., Shi, X., 1997. Neutrino production of deuterium
in supermassive stars and possible implications for deuterium detections in
Ly-limit systems. Astrophysical J., 503, 307. 

\vskip 0.1in
\noindent
Gnedin, N. Y., Ostriker, J. P., 1992.
Light element nucleosynthesis - A false clue?
Astrophys. J., 400, 1-20.

\vskip 0.1in
\noindent
Hansen, S. H., Mangano, G., Melchiorri, A., Miele, G., Pisanti, O., 2002.
Constraining neutrino physics with BBN and CMBR.
Phys. Rev. D65 023511. 

\vskip 0.1in
\noindent
Holtmann, E., Kawasaki, M., Moroi, T., 1996. Solving the
crisis in Big-Bang nucleosynthesis by the radiative decay of an exotic 
particle. Phys. Rev. Lett. 77, 3712-3715.

\vskip 0.1in
\noindent
Hu, W., Sugiyama, N. , 1994.
Thermal history constraints on the isocurvature baryon model.
Astrophys. J. 436, 456.

\vskip 0.1in
\noindent
Jedamzik, K., 2000. Lithium-6: a probe of the early Universe.
Phys. Rev. Lett. 84, 3248-3251.

\vskip 0.1in
\noindent
Jedamzik, K., Fuller, G. M., 1995. Nucleosynthesis in the
presence of primordial isocurvature baryon fluctuations. Astrophysical J.,
452, 33-61.

\vskip 0.1in
\noindent
Jedamzik, K., Fuller, G. M., 1997. Is deuterium in high 
redshift lyman limit systems primordial?. Astrophys. J. 483, 560-564. 

\vskip 0.1in
\noindent
Jedamzik, K., Rehm, J. B., 2001. Inhomogeneous Big Bang 
nucleosynthesis: upper limit on $\Omega_b$ and production of lithium,
beryllium, and boron. Phys. Rev. D64, 023510.

\vskip 0.1in
\noindent
Kang, H., Steigman, G., 1992. Cosmological constraints on neutrino
degeneracy. Nucl. Phys. B372, 494-520. 

\vskip 0.1in
\noindent
Kurki-Suonio, H., Jedamzik, K., Mathews, G. J., 1997.
Stochastic Isocurvature Baryon Fluctuations, Baryon Diffusion, 
and Primordial Nucleosynthesis. Astrophysical J. 479, 31

\vskip 0.1in
\noindent
Kurki-Suonio, H., Sihvola, E., 2001.
Inhomogeneous Big Bang nucleosynthesis and the high baryon density suggested 
by Boomerang and MAXIMA.
Phys. Rev. D63, 083508.

\vskip 0.1in
\noindent
Lindley, D. 1980. Primordial black holes and the deuterium abundance.
Mon. Not. R. Astr. Soc. 193, 593-601.

\vskip 0.1in
\noindent
Malaney, R. A., Mathews, G. J., 1993. Probing the early Universe: a
review of primordial nucleosynthesis beyond the standard model.
Phys. Rep. 219, 145-229.

\vskip 0.1in
\noindent
Marigo, P., Girardi, L., Chiosi, C., Wood, P. R., 2001.
Zero-metallicity stars. I. evolution at constant mass.
Astron. and Astrophys., 371, 152-173.

\vskip 0.1in
\noindent
O'Meara, J. M., Tytler, D., Kirkman, D., Suzuki, N., Prochaska, J. X., 
Lubin, D., Wolfe, A. M. 2001.
The deuterium to hydrogen abundance ratio towards a fourth QSO: HS0105+1619.
Astrophys. J. 552, 718-730.

\vskip 0.1in
\noindent
Orito, M., Kajino, T., Mathews, G. J., Boyd, R. N., 2000.
Neutrino degeneracy and decoupling: new limits from primordial nucleosynthesis
and the cosmic microwave background. submitted to Astrophys. J., astro-ph/0005446.

\vskip 0.1in
\noindent
Protheroe, R. J., Stanev, T., Berezinsky, V. S., 1995.
Electromagnetic Cascades and Cascade Nucleosynthesis in the Early Universe.
Phys. Rev. D51 4134-4144.

\vskip 0.1in
\noindent
Ryan, S.G., Beers, T. C., Olive, K. A., Fields, B. D., Norris, J. E., 2000.
Primordial lithium and Big Bang nucleosynthesis.
Astrophys. J. 530, L57.

\vskip 0.1in
\noindent
Sarkar, S. 1996.
Big Bang nucleosynthesis and physics beyond the Standard Model.
Rept. Prog. Phys. 59, 1493-1610.

\vskip 0.1in
\noindent
Sigl, G., Jedamzik, K., Schramm, D. N., Berezinsky, V. S.,
1996. Helium photodisintegration and nucleosynthesis: implications for 
topological defects, high energy cosmic rays, and massive black holes.
Phys. Rev. D52, 6682-6693.

\vskip 0.1in
\noindent
Webb, J. K., Carswell, R. F., Lanzetta, K. M., Ferlet, R., 
Lemoine, M., Vidal-Madjar A., Bowen, D. V., 1997. A high deuterium
abundance at z=0.7. Nature 388, 250.

\vskip 0.1in
\noindent
Woosley, S. E., 1977. 
Neutrino-induced nucleosynthesis and deuterium. Nature, 269, 42-44.

%\end{thebibliography}
\end{document}